# Synthesis and transport properties of epitaxial Bi (0001) films on GaAs (111) substrates

Jagannath Jena[1], Eugene Ark[1], Justin Wood[1], Junyi Yang[1], John Pearson[1], Hanu Arava[1], Daniel Rosenmann[3], U. Welp[1], J. S. Jiang[1], Deshun Hong[1], and Anand Bhattacharya[1]

[1]Material Science Division, Argonne National Laboratory, Lemont, Illinois 60439, USA

[2]Department of Physics, Case Western Reserve University, Cleveland, Ohio 44106, USA

[3]Center for Nanoscale Materials, Argonne National Laboratory, Lemont, Illinois 60439, USA

In recent decades, the growth of ultrathin epitaxial bismuth (Bi) films on various substrates has garnered interest due to their topological, thermoelectric, and even ferroelectric properties. We report upon the growth and transport properties of epitaxial Bi (0001) films in the thickness range of 5-32 nm deposited directly on GaAs (111) substrates, without a buffer layer. The quality of Bi films are found to depend on conditions for substrate treatment using $Ar^+$ ion-milling and annealing. Substrates milled at low ion beam currents display poor surface reconstruction after annealing, which hinders the growth of high-quality films. In contrast, substrates milled under optimized conditions led to reconstructed surfaces upon annealing, resulting in epitaxial Bi films with predominantly single-domain orientation. Although epitaxial films formed in both cases, transport measurements indicated significantly higher conductivity for films grown on optimally treated substrates. Measurements at low temperatures suggest that the transport properties are dominated by a surface state. Magneto-transport measurements suggest that conductivity and mobility improve progressively with increasing film thickness. Remarkably, our 5 nm Bi film demonstrated resilience to ambient conditions without oxidation, with transport properties similar to those of bilayer Bi films grown by confinement heteroepitaxy in other studies. These results provide a robust methodology for growing high-quality epitaxial Bi films on GaAs (111) and offer insights into their unique transport properties.

**Introduction**

Elemental bismuth (Bi) is a unique semi-metal with notable characteristics such as a very low carrier density, small effective mass, and long mean free paths for charge carriers. Bi has a rhombohedral unit-cell and its electronic structure consists of three small ellipsoidal electron pockets at the *L*-points and one small hole pocket at the *T*-point in the Brillouin zone[1,2]. However, it is known that in Bi films in the ultra-thin limit, the reduced dimensionality gaps out the bulk states due to quantum confinement[3], and entirely distinct structural and electronic properties can emerge, including the quantum spin Hall effect[4], ferroelectricity[5], and the potential for ideal 2D topological insulators[6]. Therefore, there is increasing interest in the development of high-quality crystalline thin films of bismuth, which has led to efforts to grow them on various substrates, including sapphire[7], pyrolytic graphite[8], InSb[9], glass[10], mica[11], Ge (111)[12], and different reconstructed surfaces of Si (111)[11,13,14]. Of particular interest are thin films of Bi (0001) [hexagonal indexing, equivalent to (111) in rhombohedral indexing Figure 2a], whose spin-split surface states have been studied extensively for their fundamental properties[15-24], and which in the limit of a single Bi bilayer[25] are predicted to host quantum spin Hall edge states with a large bulk gap[26]. In thicker films and bulk single crystals, the two branches of the spin-split surface states of Bi (0001) lead to a hexagonal electron pocket around the $\bar{\Gamma}$- point, six hole pockets along the $\bar{\Gamma}$-$\bar{M}$ direction, and six needle-like electron pockets near the $\bar{M}$ point. For films with thickness in the range of few nm, angle-resolved photoemission spectroscopy (ARPES) measurements have shown that bulk states are progressively gapped away from the Fermi surface with reduced thickness[21]. Furthermore, the degeneracy between the spin-split states near the $\bar{\text{M}}$-point, or lack thereof, has been the subject of several studies, and the possible topological nature of these states is still debated[20,23]. However, achieving the correct structural phase in large area epitaxial Bi (0001) films in the 1-50 nm thickness range on insulating substrates remains challenging, as the growth kinetics often do not favor the desired phase, leading instead to island-like morphology and the presence of secondary phases like Bi crystallites in the 110 orientation[27].

In this study, we present a technique for growing high-quality epitaxial Bi films on GaAs (111) substrate at room temperature using molecular beam epitaxy. Our results demonstrate that thorough Argon ion milling to remove native oxide layers combined with high-temperature annealing of the substrate is essential for achieving the correct phase of the epitaxial Bi films. When the substrate's surface reconstruction is carefully optimized, the grown Bi films exhibit smooth surfaces with large, preferentially unidirectional domains, leading to high-quality epitaxial structure. These films show significantly enhanced mobility and higher magnetoresistance compared to those grown on poorly reconstructed substrates. For Bi films with thickness in the 15-32 nm range grown on optimized substrates, a high mobility electron pocket dominates transport properties at low temperatures, while for films in the 5-10 nm thickness range, a dominant hole-type carrier contribution is observed.

**Experimental detail, results and discussion**

Crystalline elemental films play an important role in modern technology[28-32] as well as in fundamental explorations[33,34]. On an appropriate substrate, a rule of thumb for growth of crystalline films is that the growth temperature (in K) is about 0.25-0.5 of the melting point of the metal, though there are exceptions[35]. In this study, we reported the layer-by-layer growth of elemental Bi films, at room temperature, which crystallizes in the trigonal phase on top of a GaAs (111) substrate. The detailed substrate preparation and growth process are outlined below.

The treatment of the substrate prior to Bi growth is particularly important, as it significantly enhances the magneto-transport properties. The GaAs (111) substrate is first sonicated in acetone and then in isopropanol, each for 5 minutes, before being placed in the ion milling chamber to remove the native oxide layer and achieve a smooth surface. Two substrates (S#1 and S#2) were milled separately, using different ion beam currents in a background pressure of Argon gas of approximately $1.4 \times 10^{-4}$ Torr. The $Ar^+$ ion beam is incident at approximately 45°, with $V_{Beam}$ = 600 V and $V_{Acc}$ = 100 V. Sample S#1 was first milled with an ion beam current of 16 mA for 90 seconds. After this initial milling step, the ion beam was turned off for 90 seconds to prevent excess heat buildup on the surface. Milling was then resumed for an additional 90 seconds. Sample S#2 followed the same protocol but with higher ion beam currents of 17 mA for the first 90 s and 18 mA for the second 90 s period.

Once the ion milling process is complete, we immediately transferred the milled GaAs (111) substrate to a vacuum chamber ('prep' chamber, connected to the MBE chamber) with a base pressure of $5 \times 10^{-9}$ Torr to minimize exposure to air. We observed that if the milled substrate was exposed to air for more than 15 minutes, it resulted in poor crystalline quality with high surface roughness or amorphous Bi grown film, likely due to the substrate's high sensitivity to oxidation. To mitigate this, the next step involved degassing the substrate at 220°C for 45 minutes to 1 hour in vacuum, which helps remove contaminants such as water, hydrocarbons, and oxygen, creating a clean surface. Following degassing, the substrate was transferred in vacuum to the MBE chamber (base pressure $< 5 \times 10^{-10}$ Torr) and annealed at 600°C for 5-6 minutes to eliminate any remaining organic residues, smoothen the surface, and promote the desired surface reconstruction. This process is believed to enhance nucleation and adhesion, leading to high-quality thin films with improved properties. During the annealing, we monitored the reflection high-energy electron diffraction (RHEED) pattern of the substrate using a 10 kV acceleration voltage. For S#2, we observed the first signs of RHEED streaks around 500°C, with the majority of streaks becoming visible at 550°C. At 600°C, after maintaining the temperature for 5-6 minutes, half-order streaks were observed, consistent with a 2 x 2 surface reconstruction (see Fig. 1). However, for S#1, no RHEED streaks were detected up to 550°C. Upon reaching 600°C, only the integer order streaks appeared after a 5-7 minute wait, as shown in Fig. 1(a), with no visible half-order features. Further annealing at 600°C led to a decrease in the intensity of these streaks, and increasing the annealing temperature did not result in any improvement.

After annealing, the substrate is cooled down rapidly to room temperature by shutting power to the substrate heater. The substrate was typically allowed to cool further overnight with the manipulator held at 19°C. Bi films are then deposited in a typical base pressure < 1.5 x $10^{-10}$ Torr in the chamber. The deposition rate, monitored by a quartz crystal microbalance (QCM), was approximately 0.013 Å/sec for all films in this study.

The RHEED pattern during Bi (0001) film growth is shown in Figure 1. For S#1, 3D spotty RHEED patches were observed for the first 5-8 nm of deposition (Fig. 1(b)), indicating island-like growth. These islands are supposed to be Bi (10-12)[12] (Bi (110) in rhombohedral indices) as we have also observed in x-ray scans of films in this range. As the deposition continues and the thickness increases, these islands coalesce, eventually leading to the growth of an epitaxial Bi (0001) film, where the peaks corresponding to the (10-12) oriented phase are no longer visible in x-ray diffraction. After the deposition of 24 nm Bi, the RHEED intensity was weak as shown in Fig.1c, requiring a higher emission current (-2.1 µA) for better visibility as shown in Fig. 1(d), indicating a relatively poor surface quality of the Bi film. In contrast, S#2 exhibits clear 2D RHEED streak patterns at a much lower Bi thickness of around 7 nm, even with significantly lower emission current (-0.6 µA). As the Bi deposition continues, the intensity of the streaky RHEED pattern increases, and by 24 nm of film growth sharp 2D diffraction spots are observed indicating nearly atomically smooth crystalline surfaces (Fig. 1(f-h). Further examination of the surface quality was conducted using atomic force microscopy (AFM) images, shown in Fig. 2. The triangular patterns in the AFM images confirmed the Bi (0001) with trigonal axis is perpendicular to the film surface. Notably, in S#1, the domains were randomly oriented (Fig. 2b), while in S#2, the domains predominantly follow a preferential direction (Fig. 2c), which can be interpreted as a local absence of rotational twin domains[24,36].

X-ray diffraction (XRD) measurements of the two films are carried out on a Bruker D8 system and the results are presented in Fig. 2d . The (0003) peaks at 22.354° and 22.440°, as well as the (0006) peaks at 45.645° and 45.752° in 2theta-scan for S#2 and S#1, and notably the (10-12) peaks are absent, consistent with a highly crystalline single-phase Bi (0001) film. These peak positions differ slightly from the bulk Bi sample values of 22.491° and 45.888°, suggesting some deviations in the film's lattice structure possibly due to the lattice mismatch between the substrate and the film. For S#1, the rocking curve for the (0003) peak has a full width at half maximum (FWHM) of $0.157^0$, which is nearly double that for S#2 of $0.098^0$, indicating a larger mosaic spread in S#1 compared to S#2, as shown in Fig. 2(d). This difference highlights the variation of crystalline quality with substrate preparation.

To further investigate the epitaxial relationship between the GaAs substrate and Bi films, we performed X-ray reflectivity (XRR) and phi-scan XRD measurements on the (01-14), (10-15) peaks of Bi and (220) peak of the GaAs substrate, as shown in Fig. 2e and 2f . The oscillations in the XRR curves of S#1 damped more rapidly than that of S#2, despite both having nearly the same thickness (see Supplementary Fig. 1). This suggests that the surface and/or interface of S#1 is relatively rougher compared to S#2. The in-plane phi-scan of the (01-14) Bi peak reveals a total of twelve peaks, with six strong peaks and six weaker peaks. For S#2, the intensity of the weaker peaks ~ $10^{-2}$ of the stronger peak, while for S#1 this ratio is $10^{-1}$, suggesting the presence of four sets of twin planes. Notably, the GaAs peaks are offset by ~ 30° in φ from peaks corresponding to the majority Bi (0001) domains, and are thus at approximately the same azimuthal orientation (within 0.3°) as the minority Bi domains. Additionally, the intensity of the Bi peaks in S#2 is approximately 1.5 times stronger than in S#1, signifying a better in-plane domain alignment in S#2 compared to S#1.

The Bi films were lithographically patterned into six-probe Hall bar devices for magneto-transport measurements. We first discuss our results from four samples, S#1and S#2 with thickness 24 nm of Bi (0001), and also two other films of thickness 16 nm and 32 nm both prepared under conditions identical to S#2. The Hall bar device width was 200 μm, and the distance between the voltage tabs was approximately 700 μm (Fig. 3(a)), and the distance between the current leads was approximately 2 mm. Pt/Au contact pads were deposited for wire bonding to the leads. The magnetic field was applied perpendicular to the film plane for all measurements. The longitudinal resistance and transverse Hall signals were measured by sweeping the magnetic field from +14 T to -14 T, and to eliminate the effect of the magnetic field on the longitudinal component of the resistivity contributing to the measured Hall voltage, we anti-symmetrized the Hall signal.

Shown in Fig. 3(d) is the change in sheet resistance with decreasing temperature for S#1 and S#2 samples. The increase in resistance with decreasing temperature for both the samples suggests a transition from semi-metallic to semiconductor-like behavior, driven by the bulk of the thin film becoming insulating while the surface remains predominantly conducting. This contrasts with bulk Bi crystals, which exhibit metallic behavior with resistivity decreasing with decreasing temperature[37-39]. It is worth noting that the resistance in S#1 is higher compared to S#2. Fig. 3(b, c) and 3(e, f) showed the longitudinal magnetoresistance (MR) and transverse conductivity for S#1 and S#2 at different temperatures. The MR and transverse conductivity are defined as:

$$MR = \frac{\rho_{xx}(B) - \rho_{xx}(0)}{\rho_{xx}(0)} \qquad \text{Eq. 1}$$

where $\rho_{xx}(B)$ and $\rho_{xx}(0)$ are the resistivity with and without magnetic field B applied perpendicular to the film plane at a given temperature. The resistivity is defined as $\rho_{xx} = R_{xx} W \cdot t / L$, where $W$ (200 μm) and $L$ (700 μm) are the width and length of the Hall bar, respectively, and $t$ is the thickness of the Bi films. In a simplified 2-band model, the transverse conductivity is given by:

$$\sigma_{xy} = -\frac{\rho_{xy}}{\rho_{xy}^2 + \rho_{xx}^2} = eB \left[ \frac{n_h \mu_h^2}{1 + \mu_h^2 B^2} - \frac{n_e \mu_e^2}{1 + \mu_e^2 B^2} \right] \qquad \text{Eq. 2}$$

where $\rho_{xy}$ (= $R_{xy} \cdot t$) is the transverse resistivity and $n_h$ ($n_e$) and $\mu_h$ ($\mu_e$) are the hole (electron) carrier density and mobility, respectively.

While we do not have an adequate model to describe the longitudinal MR, a few points are worth mentioning. In S#2, there was a relatively sharp rise in MR up to approximately ±4T, followed by a gradual increase, reaching 18.7% at ±14T. In contrast, MR in S#1 showed weaker response to the magnetic field, with its value peaking up to 8.35% at ±14 T (Fig. 3(c)). The significant scattering in S#1, as evident from lower MR and higher values of resistance vs temperature, suggests that it has lower mobilities compared to S#2. To explore this further, we plotted the transverse conductivity ($\sigma_{xy}$) of both samples, as shown in Fig. 3(c) and 3(f). The $\sigma_{xy}$ of S#2 at 5 K exhibited a highly nonlinear inverted S-shaped curve with negative curvature and a sharp min/max at ±1.3 T. Beyond these extrema, $\sigma_{xy}$ decreases in magnitude eventually crossing zero and changing sign at 9.6 T. On increasing the field further, $\sigma_{xy}$ increases linearly in magnitude, as shown by the blue curve, up to 14 T. In contrast, S#1 at 5 K showed positive curvature in the low-field region, and $\sigma_{xy}$ remained positive (negative) throughout the entire accessible positive (negative) magnetic field range. The positive slopes of the $\sigma_{xy}$ curves at high magnetic fields indicate the presence of hole-type carriers in both samples. As we shall see below, the negative slope at lower fields (from 0T to 1.3T) and the inverted S-shape observed in S#2 originates from high mobility electrons.

When both electron and hole types co-exist in a semimetal system, a two-band model may be used to fit the transverse conductivity curves[40]. By using the expression from Eq. 2, it was found that at 5K, S#2 has much higher electron mobility and electron concentrations ($\mu_e = 6642$ cm$^2$V$^{-1}$s$^{-1}$; $n_e = 4.99 \times 10^{11}$ cm$^{-2}$) than S#1 ($\mu_e = 2942$ cm$^2$V$^{-1}$s$^{-1}$; $n_e = 8.4 \times 10^{10}$ cm$^{-2}$). We note that the holes have relatively low mobility and give a quasi-linear contribution $\sigma_{xy} \sim (n_h \mu_h^2)eB$ at the magnetic fields in this study, where the slope does not allow us to uniquely determine $\mu_h$ and $n_h$. The presence of multidomain structures, larger defects, and grain boundaries in S#1 likely disrupted charge transport, particularly at low temperatures, leading to lower conductivity compared to S#2, which offered more efficient electron transport due to fewer scattering centers.

We note that for all samples, $n_e$ increases with temperature. This suggests that the very low carrier density in the high-mobility electron pocket that dominates $\sigma_{xy}$ at low fields can change significantly due to thermal activation. We also note that S#1 also has a much lower $n_e$ than S#2 at the lowest temperatures. This implies that the position of the Fermi level in this very small electron pocket is sensitive to conditions under which the substrate is prepared. At higher temperatures, electron-phonon scattering becomes more significant than scattering from grain boundaries and defects, leading to stronger temperature dependence. Furthermore, thermal excitation leads to higher carrier densities, particularly for electrons in the high mobility pocket, resulting in higher conductivity, a pronounced 'inverted-S' shape $\sigma_{xy}$ at low fields, and higher magnetoresistance in both samples. We note that S#1 remains distinct from S#2 at all temperatures - the maximum conductivity and magnetoresistance reached up to 480 $\Omega^{-1}\cdot$cm$^{-1}$ and 80% in S#2, compared to 450 $\Omega^{-1}\cdot$cm$^{-1}$ and 65% in S#1. We also obtained carrier concentration and the mobility values of the electrons at various temperatures using the two-band model fitting of the conductivity curves (Fig. 4 and Table 1).

To investigate how conductivity and magnetoresistance (MR) vary with thickness, we grew 16 nm and 32 nm Bi films in conditions identical to S#2. The R–T curves of the two films, shown in Fig. 3(g) and Fig. 3(j), reveal that the resistance of the 16 nm Bi film is more than twice that of the 32 nm Bi film. The 32 nm film exhibited very large MR (around 115%) and conductivity (647 $\Omega^{-1}\cdot$cm$^{-1}$), while the 16 nm film showed significantly lower values of 35% and 360 $\Omega^{-1}\cdot$cm$^{-1}$, respectively, indicating that thinner films have notably reduced transport properties. In bulk Bi crystals, magnetoresistance (MR) typically exhibits its highest values at low temperatures[41-43] and gradually decreases with increasing temperature. In contrast, thin Bi films display an opposite trend, where MR increases with temperature and reaches a maximum around 200 K for all samples (see Fig. 3). At room temperature, MR becomes nearly linear, with a slightly reduced value compared to the peak at 200 K. For the 16 nm, 24 nm (S#2), and 32 nm films, at low temperatures (5K) the MR rises sharply at low magnetic fields and then more gradually at higher fields. An exception is observed in the hole-carrier-dominated 24 nm S#1 film, which shows a different field dependence at 5 K, indicating a different transport mechanism. A detailed analysis of this unusual temperature- and sample-quality-dependent MR behavior is beyond the scope of the present study. We also compare the $\sigma_{xy}$ vs B measurements for different thicknesses at low temperature (Fig. 4a) and also the temperature dependence of $n_e$ and $\mu_e$. We note that the mobility in the 32 nm Bi film is the highest among all the thickness investigated, whereas the 16 nm film has the lowest mobility, while still being greater than the mobility for the 24 nm S#1 film. Table 1 summarizes the mobilities and carrier concentrations of all the films studied.

As previously mentioned, through optimized substrate treatment and annealing conditions, we grew epitaxial Bi (0001) films down to thickness of approximately 5 nm. These films exhibit excellent stability in ambient conditions without oxidation, which enabled us to perform various ex-situ measurements, including AFM, XRD, XRR, in-plane phi scans, and transport measurements on patterned devices (See Fig. 5 and supplementary Fig. 2). AFM analysis revealed a smooth surface with an estimated root mean square (RMS) roughness of 0.41 nm. The (01-14) phi scan showed peaks at similar positions to those observed in thicker films, albeit with significantly weaker intensities. Additionally, resistivity measurements indicated an increase with decreasing temperature, suggesting an insulating behavior in this ultra-thin film. Interestingly, similar resistivity trends have been observed in ultra-thin bismuth layers (~0.39 nm) grown between a 6H-SiC substrate and epitaxial graphene via confinement heteroepitaxy[25]. Furthermore, we note an increase in resistance upon cooling to the lowest temperatures, starting at around 4.5

K. This behavior, also observed in thin films of topological materials[44], could be due to electron-electron interactions. Importantly, Hall measurements revealed sublinear curves, similar to those observed in the 24 nm S#1 film at low temperatures. As the temperature increased, the Hall curves continued to display linear behavior at higher temperatures, in contrast to the S#1 sample, which exhibited inverted S-shaped curves due to the presence of both hole and electron carriers (supplementary Fig.3a). This suggests that in the ultra-thin 5 nm Bi film, the Hall signal is primarily contributed by hole-like carriers. The high mobility electron pocket visible in thicker films is gapped out, and does not get populated even at higher temperatures. It is known from ARPES measurements and from theoretical calculations that the high-mobility electron pockets near the $\bar{M}$-points vanish at very low thicknesses due to the hybridization of surface states with bulk states near the M-point, and confinement effects that move these bulk away from the Fermi level[45]. Consequently, the transport properties become dominated by the relatively low mobility hole pockets that lie along the $\bar{\Gamma}$-$\bar{M}$ direction, as observed in the 5 nm Bi film. We also observed a cusp-like behavior, characteristic of the weak antilocalization (WAL) effect, in the low-field region near zero to the $R_{xx}$ curve (red). This effect is typically observed at low temperatures in thin films of two-dimensional systems with strong spin-orbit coupling (SOC), due to weak anti-localization[46]. Similar cusp-like behavior is also present in other thicker samples but is less pronounced in the 24 nm (S#2) and 32 nm films and tends to vanish with increasing temperatures (see supplementary Fig.3b).

**Conclusions:**

In conclusion, for ultrathin epitaxial Bi films grown on GaAs (111) substrates, we found that argon ion milling, annealing, and growth conditions significantly affect film growth and electrical transport properties. For films grown on substrates prepared under conditions where the 2 x 2 reconstruction was not as evident, the films showed lower mobilities. In contrast, films on substrates with optimized milling and annealing conditions where the 2 x 2 reconstruction was evident exhibit smoother surfaces, larger unidirectional domains, higher conductivity, and larger positive MR in perpendicular fields, and show strong contributions from high-mobility electrons at low temperatures. To deepen our understanding, we grew films of different thicknesses following the later process and found that, with increasing thickness, the MR, conductivity, and mobilities all increased. The conductivity of the thicker films exhibited inverted S-shaped curves, consisting of a negative slope at low fields and a positive slope at high fields, indicating contributions from both hole and electron pockets. In contrast, the 5 nm Bi films displayed a negative slope and a sublinear curve in the Hall resistance, suggesting that transport is primarily dominated by holes. Our findings provide a clear procedure for growing high-quality Bi thin films on GaAs (111) substrates with potential applications in spintronics and in fundamental studies of the spin textured surface states of Bi (0001).

**Acknowledgements**: The synthesis and characterization of samples presented in this study, and their phenomenological modeling was supported by the Center for the Advancement of Topological Semimetals (CATS), an Energy Frontier Research Center funded by the U.S. Department of Energy (DOE) Office of Science (SC), Office of Basic Energy Sciences (BES), through the Ames National Laboratory under contract DE-AC02-07CH11358. The calculations of the surface-state electronic structure and electron mobility in Bi thin films were carried out by S. Ambhire and S. S.-L. Zhang, with support from the College of Arts and Sciences at CWRU. E.A.,

J.W., and J.Y. assisted in this research and were supported by the Materials Science and Engineering Division, Office of Basic Energy Sciences, U.S. Department of Energy. Work performed at the Center for Nanoscale Materials, a U.S. Department of Energy Office of Science User Facility, was supported by the U.S. DOE, Office of Basic Energy Sciences, under Contract No. DE-AC02-06CH11357.

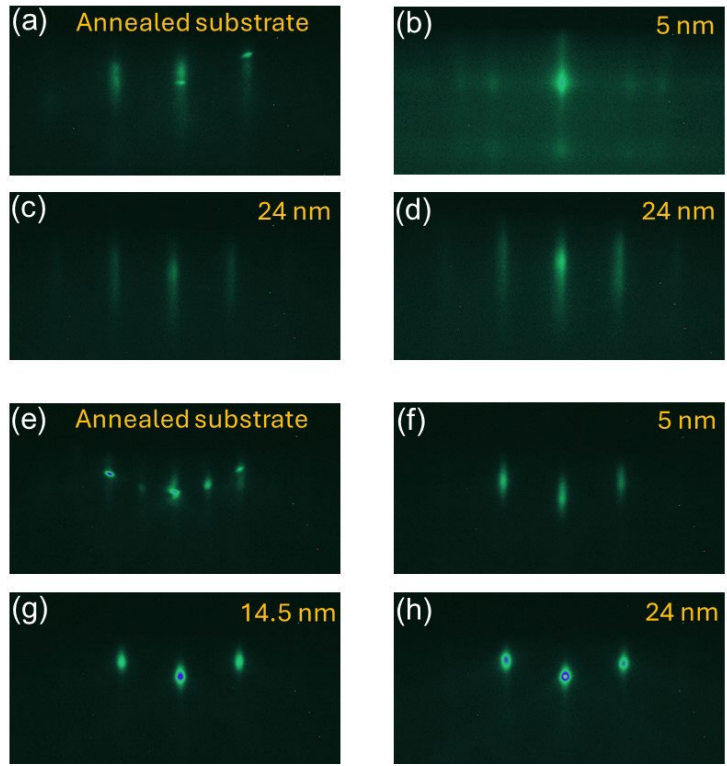

**Figure 1:** (a) RHEED pattern of the GaAs (111) substrate for S#1 taken along the $(11\bar{2})$ azimuth after annealing. Prior to annealing, the substrate was ion milled with an Ar$^+$ ion beam current of 16 mA. The ion milling process was performed twice, each time for 90 s with a 90 s gap between the milling sessions. (b) A 3D spotty RHEED pattern indicating island growth of Bi in the initial stages of growth. (c, d) RHEED patterns of the 26 nm epitaxial Bi film at emission currents of -0.9 μA and -2.1 μA, respectively. (e) RHEED pattern of the GaAs (111) substrate for S#2 taken along the $(11\bar{2})$ azimuth after annealing, showing evidence for a 2x2 reconstruction. Prior to annealing, the substrate was ion milled with ion beam currents of 17 mA and 18 mA for 90 s each, with a 90 s gap between the milling sessions. (f-h) In this case, epitaxial Bi film started to grow in an atomic layer-by-layer manner from an earlier stage, showing 2D RHEED patterns. The RHEED images were captured at different Bi deposition thicknesses at a lower emission current of -0.6 μA.

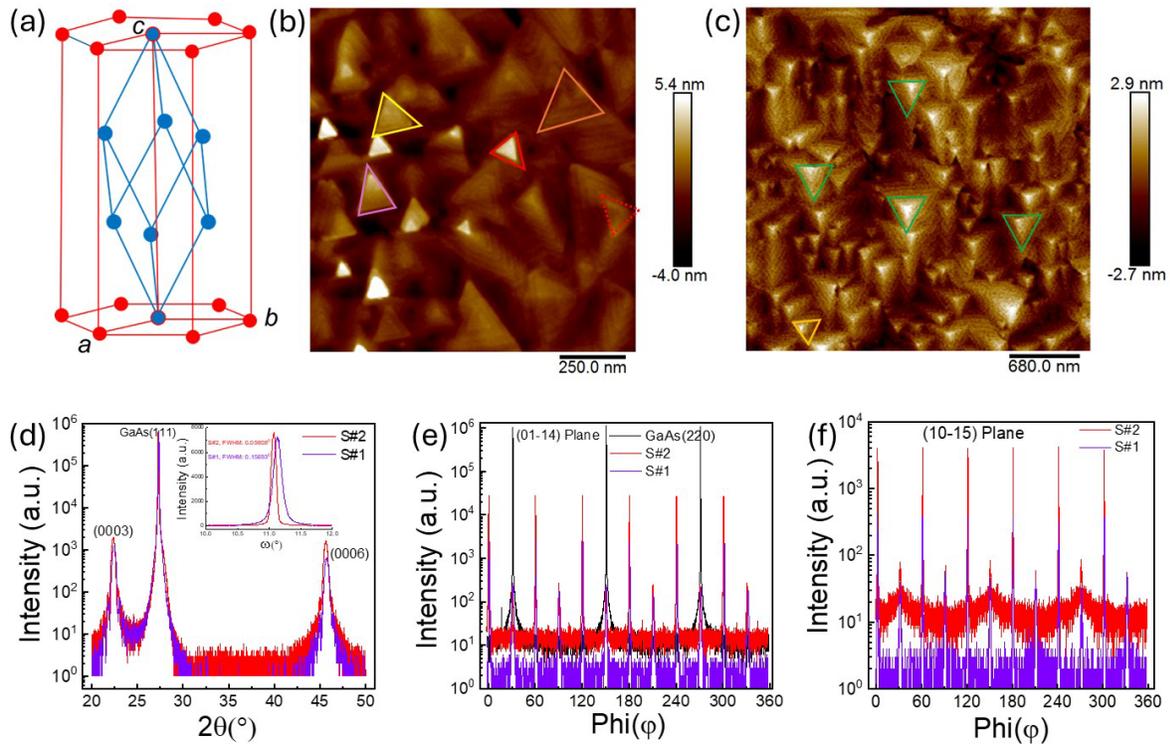

**Figure 2**: (a) Schematic diagram of rhombohedral Bi lattice (blue lines) in a hexagonal lattice (red lines). Not all atoms are shown in the schematic. (b) AFM image of sample S#1. (c) AFM image of sample S#2. Different triangular domains are marked with different colors. (d) Both Bi films peaks are denoted by (0003) and (0006) peaks of both the samples in the 2theta-theta XRD measurements. (e, f) Phi-scan of the tilted (01-14) and (10-15) planes of the S#1(purple) and S#2 (red) samples. The (220) plane of the GaAs substrate shows the three-fold symmetry (black). The weaker intensity observed for the (10-15) plane of Bi compared to the (01-14) plane is primarily attributed to a smaller structure factor, which reduces the diffracted intensity despite similar measurement geometry.

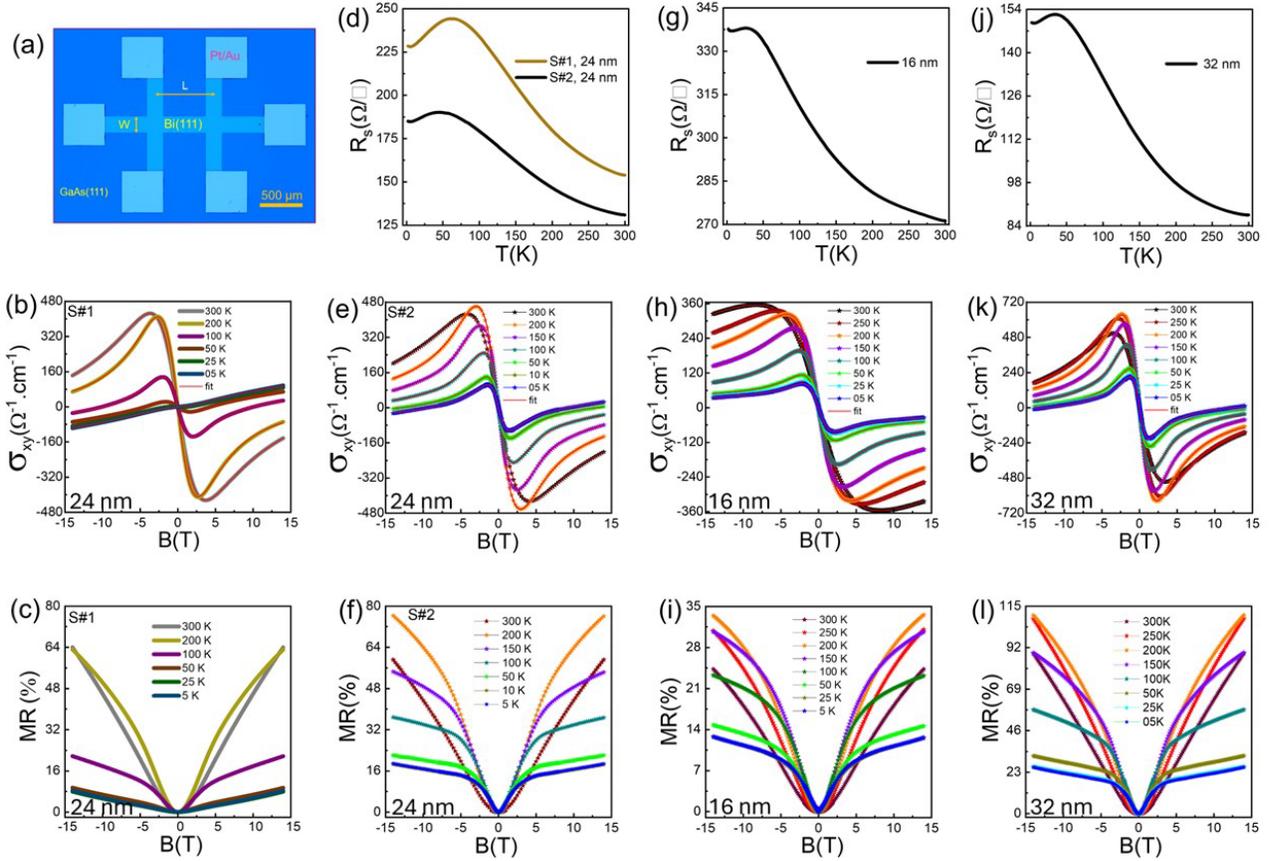

**Figure 3:** (a) Lithographically patterned Hall bar device. W and L represent the width and length of the Hall bar, respectively. The scale bar corresponds to 500 μm. (b, c) Hall conductivity and magnetoresistance of S#1 at different temperatures (d) Sheet resistance vs temperature curves for S#1 and S#2 samples, both 24 nm thick. (e, f) Hall conductivity and magnetoresistance of S#2 at different temperatures. (g, h, i) Sheet resistance, Hall conductivity and magnetoresistance with temperature of a 16 nm Bi film. (j, k, l) Sheet resistance, Hall conductivity and magnetoresistance with temperature of a 32 nm Bi film.

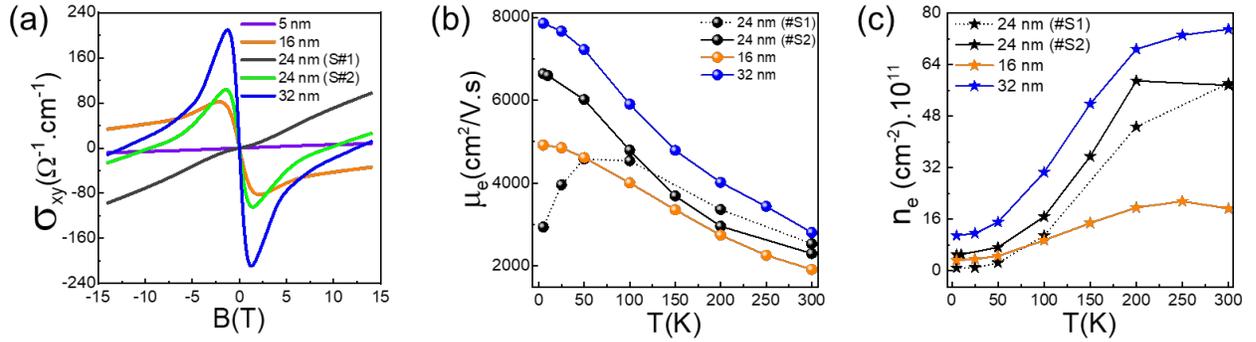

**Figure 4** (a) Hall conductivity at 5 K for a series of Bi (0001) epitaxial films with different thicknesses on GaAs(111) substrates. Sample S#1, where GaAs (111) substrate did not show evidence for a 2x2 reconstruction has a strongly suppressed Hall conductivity at low fields. S#2, with similar thickness but deposited on a substrate where the 2 x 2 reconstruction was clearly evident, shows evidence for an electron pocket with high mobility, in the Hall conductivity, along with a relatively low mobility hole pocket. (b) Mobility and (c) carrier density of electrons as a function of temperature, extracted from a two-band fit to the Hall conductivity.

| Thickness (nm) | $n_e$ (5 K) cm$^{-2}$ | $\mu_e$ (5 K) cm$^2$/Vs | $n_e$ (300 K) cm$^{-2}$ | $\mu_e$ (300 K) cm$^2$/Vs |
|---|---|---|---|---|
| 16 | $3.22 \times 10^{11}$ | 4920 | $1.93 \times 10^{12}$ | 1921 |
| 24 (S#1) | $8.4 \times 10^{10}$ | 2942 | $5.82 \times 10^{12}$ | 2538 |
| 24 (S#2) | $4.99 \times 10^{11}$ | 6642 | $5.76 \times 10^{12}$ | 2303 |
| 32 | $10.9 \times 10^{11}$ | 7855 | $7.5 \times 10^{12}$ | 2810 |

**Table 1** Properties of the high mobility electron pocket extracted from Hall conductivity.

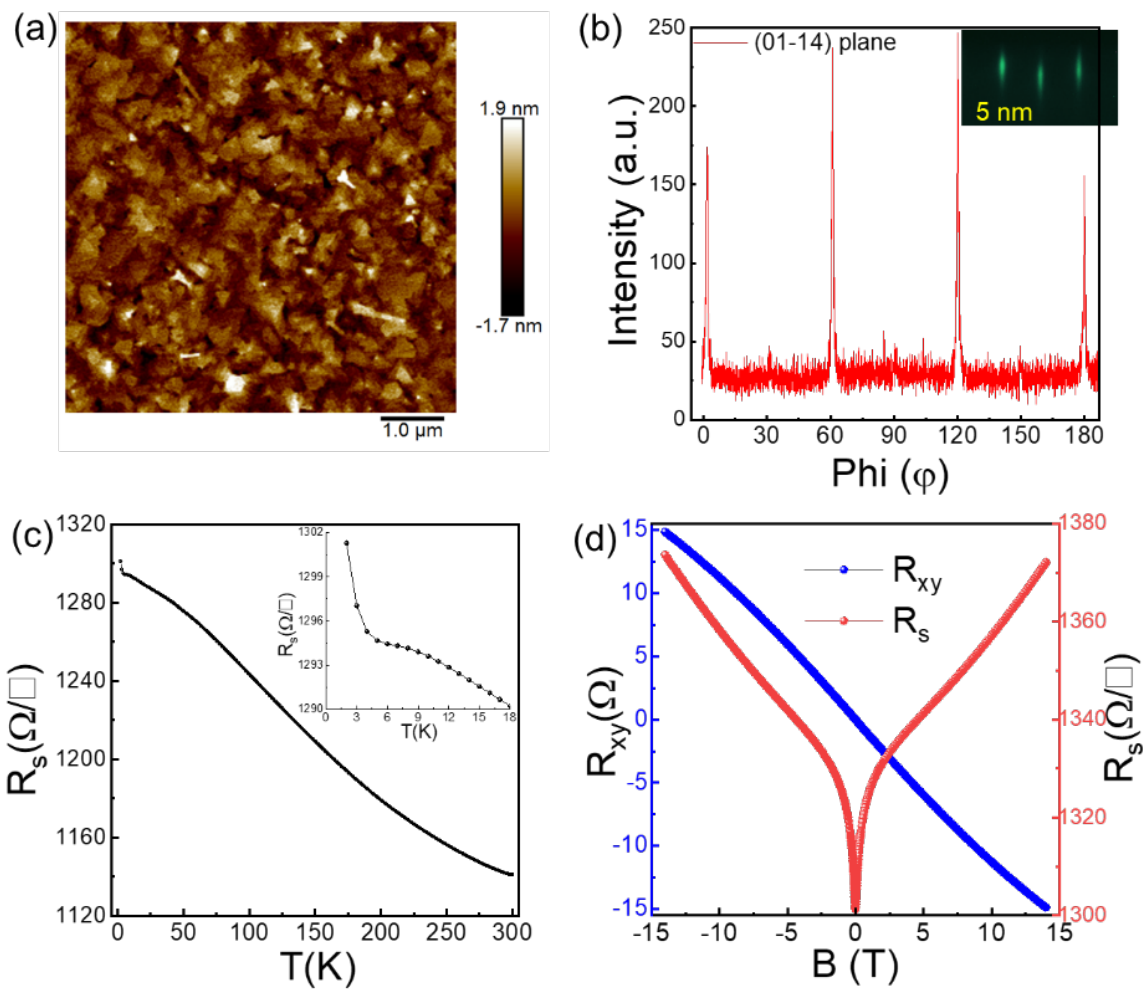

**Figure 5:** (a) AFM image of the 5 nm Bi film. (b) (01-14) phi-scan of the film. Inset showed the RHEED pattern after the growth of the film. (c) R-T curves from 300 K down to 2 K. The sharp rise of resistance from 4.5 K is shown in the inset. (d) $R_S$ (red) and $R_{xy}$ (blue) vs. magnetic field at 2 K.